\documentclass{article}

\usepackage{arxiv}

\usepackage[utf8]{inputenc} 
\usepackage[T1]{fontenc}    
\usepackage{hyperref}       
\usepackage{url}            
\usepackage{booktabs}       
\usepackage{amsfonts}       
\usepackage{nicefrac}       
\usepackage{microtype}      
\usepackage{lipsum}		
\usepackage{graphicx}
\usepackage{natbib}
\usepackage{doi}

\title{Modeling Mobile Visualization for Medical Reports of Complex Chronic Diseases}

\author{ Sankarshan Dasgupta \\
	Department of Computer Science\\
University of Dayton\\
	Dayton, OH 45469 \\
	\texttt{dasguptas2@udayton.edu} \\
	\And
	Tom Ongwere \\
	Department of Computer Science\\
	University of Dayton\\
	Dayton, OH 45469\\
	\texttt{tongwere1@udayton.edu} \\
}

\begin{document}
\maketitle
\begin{abstract}
Visualizing medical histories of patients with complex chronic diseases (e.g., discordant chronic comorbidities (DCCs)) is a challenge for patients, their healthcare providers, and their support network. DCCs are health conditions in which patients have multiple, often unrelated, chronic illnesses that may need to be addressed concurrently but may also be associated with conflicting treatment instructions. Future work targeting to reduce treatment conflicts and improve patient quality of life and care should carefully examine and visualize DCCs medical reports, symptoms, and treatment recommendations. In this study, we explore various visualization models and paradigms. We analyze how these models and paradigms are applied to visualize multifaceted medical data. We then propose a model for transforming the unstructured data into temporal slices and depict them in a single graphic model. We report how we carefully moved multifaceted DCC records into; structured data tables, visualization graphs, and various hardware devices.
\end{abstract}


\section{Introduction}

Complex chronic diseases for example Discordant Chronic Condition (DCCs) have unrelated or contradicting care and treatment plans, for example, depression, arthritis, or end-stage renal disease is discordant to type-2 diabetes, \cite{Ongwere2018-fb}. Managing DCCs often requires patients and their healthcare providers to engage in excessive self-care activities, prevent or delay symptom progression, maintain a healthy lifestyle. These may require patients to apply multiple recommendations at the same time and understand interactions among those recommendations. The modalities through which DCCs data (including symptoms and recommendations) is visualized, have a profound effect on a patient's and their healthcare providers' ability to accurately reflect and flag treatment conflict and interactions.

In current healthcare electronic records systems, visualization tools capture valuable insights from patients’ records. However, visualization continues to be a challenge when using the traditional document, these documents fail to visualize and summarize or reflect the end user experiences. In this work, we explore state-of-art visualization models and build a framework targeting successfully visualizing complex reports of patients with DCCs. Visualization is instrumental in aiding patients and doctors with better information extraction. We first report our exploration of various cross-platform and state-of-art data visualization tools and models. We then discuss their implementation using medical reports of patients with DCCs and conclude by proposing novel framework targeting art visualizing complex and often multifaceted data that patients with DCCs collect across their multiple care settings.

Many visualization tools are being implemented as part of \textit{InfoViz} for different data extraction and trends representation. However, as reported above, the use of visualization in complex experiences of patients with DCCs is a challenging task for the uniqueness of complex data reports. We proposed a novel flow model to understand the basic approach for a visualization tool to impact the long standing conundrum of reporting a DCC's experience. Framework and Models Visualizing DCC's experience and trends should actively adapt to continuous changes and dynamic representation of the complex historical data analysis. The framework identifies the following steps to intertwined; 	i) Semantics and data extraction, ii)  Visualization models and patterns, iii) Data transformation and iv) Evaluations on efficacy of the model

\begin{figure}[!ht]
  \includegraphics[scale=.1, width=\linewidth]{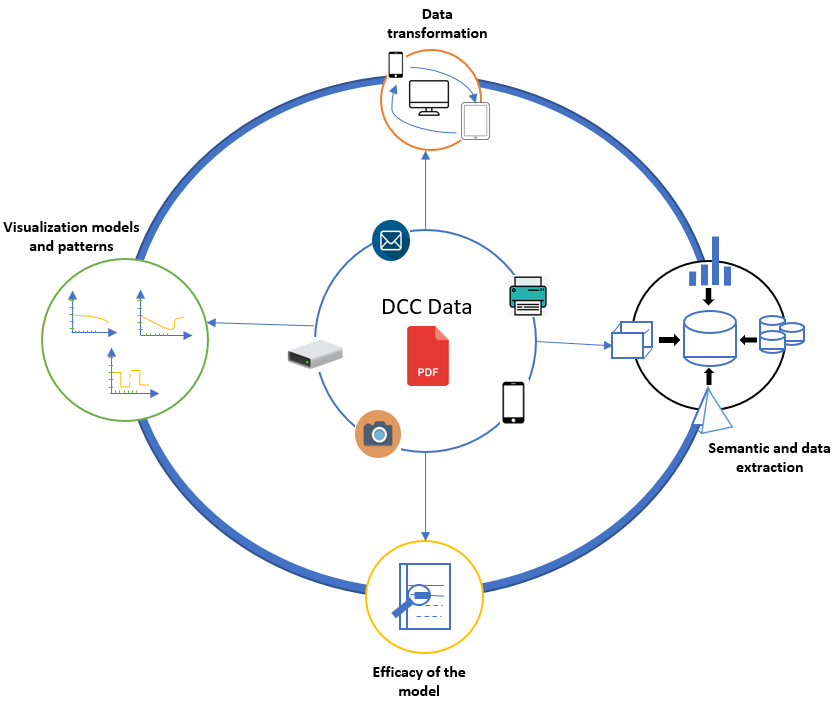}
  \caption{The architectural model is derived for discordant chronic disease data, semantics of data extraction, visualization patterns observed and efficacy mapped to it. The key feature of this model is the dependency of all task on each other making it a semi-heuristic in nature.}
  \label{fig:Model}
\end{figure}

\section{Related Work}
Because of the limited visualization work done on DCCs, this section; first highlights studies done on visualizing medical histories and trends for single chronic diseases and multiple chronic diseases. Secondly, we discuss the standard steps required to build a \textit{infoviz} model, including i) Tools Visualizing Patients’ Medical Histories and Disease Trends ii) Data collection, iii) Information visualization, iv) Visual data manipulation, and v) Evaluation of models.

\subsection{Tools Visualizing Patients' Medical Histories and Disease Trends}

Data visualization tools are used to foster patient engagement and help both patients and providers understand complex data. \cite{Lim2016-cc} explored tools that provide visualization of patient data and elicit information about important aspects of a patient’s daily life. \cite{Kyoung_Choe2017-kr} provided instructions on how t design tools to support self-reflection using visual data exploration. They include tools to reflect the frequency of treatment recommendations and prompt participants to provide potential contextual information. However, these tools may not work for patients with DCC. The complex disease interactions between type-2 diabetes and DCCs' need to be simplified before we can understand how symptoms and treatment plans may impact another symptom or treatment plan. 

\cite{Mamykina2015-wq}. recommends supporting continuous visibility and understanding, for example, creating systems where patients post their problems and suggest threads that are relevant to their posts. This support network can provide support in sense-making and there are visualizations tools designed to i) represent information that patients collect, ii) organize search results, iii) highlight important terms, and iv) annotate content and organize information. Future work should explore strategies or frameworks to visualize the discussion threads and draw the patient’s attention to areas of conflict, or to areas where the discussion pivots and may warrant further examination. There is already a study that proposes the use of a disease active score predictive model to longitudinally predict disease activity score based on C-reactive proteins \cite{Nishiguchi2014-xm}. This model uses protein as a metric to evaluate the severity of a patient's mental health. However, additional metrics are required to accurately predict the disease activity and mental state of patients with DCCs. 

\subsection{Data collection}
\label{sec:headings}

Data collection is a primary objective for any and all of the data driven models. The data from multiple reports and from different source as in: complex chronic diseases are a challenging task. Hence data collected should be logically rearranged to a structured data for any analysis to be conducted. And medical information visualization also requires the same structured data to make a meaningful representation.
\cite{Zhang2018-om} presents the process of data visualization as hierarchical task analysis rather than cognitive work analysis. Cognitive work analysis describes a system affirmatively based on constraints imposed on activities while hierarchical task analysis is more goal-specific, with plans and activity-oriented descriptions. Thus, the chronic disease data visualization approach can be influenced by abstraction work on hierarchical task analysis rooted in human-computer interaction. For visual analysis, hierarchical task abstraction can integrate individual details. Another approach that supports hierarchical task analysis is multilevel detail extraction with different attribute details.  Future work looking to achieve such kind of \textit{infoViz}, needs to consider Multivariate feature implementation. There is already some work taking that direction. For example, \cite{Nobre2019-ql} introduced a multivariate visualization technique with genealogy and environment which is an important feature for most of the analysis when described for chronic disease reports in visualization techniques. The visualization should comply to be information oriented, reflective of the data, and transparent amongst others as cited by \cite{Meyer2019-hg}

Another important construct is \cite{Whitlock2020-re} that adds creativity in complex data processing is the integration of real-time update features which could include newly extracted data as suggested. Unique visualization plots including bar charts, scatter plots, and heat maps are already being used to provide these real-time updates about data collected. Other strategies that have been used in a similar context include; synchronization of the data, data quality validation and data fusion are some of the properties which we need to address for better visualization of the data collected as mentioned by \cite{Neubert2019-jg}.
For medical histories, which are usually collected in a continuous process and high periods, current studies are using a timelines approach to represent a temporal event in a sequential format. For example, \cite{Bartolomeo_undated-zh} provides ways to visualize patient histories and symptom trends in higher time ranges and linear and radial layouts. Future work looking at similar contexts may learn from this work and gather the information for data representation. Linear range charts are instrumental in presenting transparent multiple charts compared to cyclic charts with rectangular masks when separating different visualization modules is involved. Furthermore, a plot with unique color and interpolation on the ranges and events will be helpful for future work exploring the visual representation of our chronic disease patients.

\subsection{Information visualization}
There are multiple \textit{InfoViz} tool to visualize data in an application. Our motivation arrives from \cite{Zhang2018-om} where Type 1 Diabetes patients health report such as: glucose level, which is successfully monitored and visualized using violin plot. We are trying to generalize the visualization for treatment of complex discordant chronic disease for every chronic disease patient and not limited to Type 1 Diabetes.

\cite{Ens2021-fx} have discussed the challenges encountered for research visualization space, the representation of information, designing guidelines, human perception and cognition and ethical implementation. Visualization space must be utilized for complex representation for immediate understanding. It also suggests the design should be efficient for HCI information extraction and dynamic settings in visual graphs. And most importantly to decrease working memory and cognitive burden, various methods for integrating information with the physical world. \cite{Ge2020-bc} describes an animated visualization approach called \textit{Canis}, we can select marks, partition it and apply animations to the specific mark. The uniqueness of this tool is the variety of templates offered for data visualization. There are two categorized template approaches mentioned with keenness in this research for template selection. The first is using key frames, and the second is data driven template. The latter is backed in the paper to avoid the tedious and time consuming manner of key frames. With respect to data driven template bar plots and radial plots for serving multivariate analysis are more preferred. Since this is a critical and novel scenario for medical reports representation, we have to be careful about loss of data and so \textit{Canis}  insights can help us in assisting different information visualization. We have all the derivation of visualization, but modeling of the visual charts for doctors and patients need more understanding for either users on a higher level. \cite{Schufrin2021-rd} made use of visualization for personal data on the internet stored by websites. The data exported from the internet is used to plot in a scatter plot by a tool called \textit{transparencyviz}. The plot increases the understand-ability and transparency of the user’s personal data on the internet. The tool processes the exported data for the user and implements a visualization pattern with remarkable accuracy. This is of our interest that the users can understand the gathered information which is a challenging task for medical reports. Moreover, this is a visualization of diverse format datasets and could be of helpful to all medical individuals for reading diverse formats. The stages of this tool relies on perception, comprehension and projection. Our research suggests that these stages can well be adapted in our model to build visual pattern for precise medical report visualization.

The challenge for a more general visualization is the factors such as multivariate data, time range, data charts implementation and user understanding of the visual graph. If compromised on any of the above mentioned challenges, the visualization might loose information and could be fatal in ongoing treatments. To address this concern, we have gone through intensive survey and found out many research which were performed in acknowledging the above challenges in visualization for different input data. The information we  gather from other research are discussed in the next section. Most of these work performed are on different inputs other than medical reports but could well be utilized for our research work on medical reports. The visualization are driven through multiple application for user accessibility but we have identified that most accessible and portable device is our smartphones. Thus, we will limit our visual charts implementation to mobile phones for now.

\subsection{Visual data manipulation}
The approach for designing a mobile visualization is a coercive approach with limitations in display, storage and computational power. \cite{Whitlock2020-re} gives us a comprehensive information about the difficult task of mobile visualization.  Though this approach was made for field data analysis, we found the work very meticulous for our implementation. The raw data from reports are structured to data tables using data transformation. Data transformation is an integral and critical part before information visualization. The data can be misinterpreted, lost or falsely manipulated during this task. For this reason, synchronization of the data, data quality validation and data fusion are some of the properties which we need to address for transparent visualization of medical reports.

\textit{MobileVizFixer} \cite{Wu2021-kq} is a tool to convert all the web app visualization into mobile friendly visualization. Due to lateral view, and smaller screen size affecting the visualization in a mobile device which results in distorted views, hidden view ports and unreadable information. The tool cited is used to accommodate and interpret the web based visualization into a mobile phone for better understanding. This tool works on the Markov decision process of supervised learning to find the best possible viewing of the data visualization. It consist of automatic extraction of the events is proposed from Scalable Vector Graphics (SVG) based models to mobile friendly visualizations. The tool works on its own to understand the data and interpret into a more visual friendly content with decision rules over a hybrid system. The policy based rules are implemented on the system to make it more flexible to real world scenarios. The transparency on this model also helps in explainable features for the mobile application. Many issues on adaptation such as out of view charts, blank spaces and unreadable text are acknowledged and fixed in this research. We are influenced by the technique and we think it will be helpful to integrate the design ideas onto our charts of medical reports for streamlining the visualization.

Another research by \cite{Brehmer2020-ui} derives a previous web based visualization into a mobile phone centric application. The main visualization used is scatter plot and comparison discussed between animated and small multiple visual windows. The paper mainly focuses on the participants' interaction with the set of survey responses and deriving the result of time range on the delay of interaction by the participants to conclude the user friendliness of visualization. This approach can be used for our research to make an efficient model so that all the important details from a single user interface can be extracted by the medical personal from the visual graphs of the patients. These could give us a base for our new visualization tool which could help in medical reports for chronic diseases. To verify the performance of our tool, we could use a previously researched survey papers for assistance. We will also try to implement a cross platform visualization algorithm which can be projected on bigger screen and projectors without loosing pixel or the aspect ratio.

\subsection{Evaluation of models}
Complex diseases involves discordant chronic comorbidity \cite{Sevick2007-xh} that require the discrete attention from multiple health care providers. The information is spread across multiple reports by health care professionals and is very complicated for data-management. With data linking from multiple reports, it increases information processing complexity for health care professionals and patients alike. Multiple chronic diseases are likely to require the attention to multivariate details and behavioral changes for long time spans for appropriate treatment. This could be a motivation for future research on this conundrum and build a visualization tool to assist such complex architecture for doctors and patients.

To build a \textit{infoviz} we started with a reference point on the previous paper discussed in the above subsections. \textit{IdmViz} by \cite{Zhang2018-om} has given us the way forward for its visualization on Type 1 Diabetes patient reports. In more general case, we could take into consideration the timelines and multivariate nature of the tool. The dynamic nature of \cite{Whitlock2020-re} on immersive analysis tool for the field study discussed in the previous section supports our tool design for real time visualization. With more analysis we have found that \cite{Brehmer2020-ui} could be ideal implementation for our DCC data reports which supports higher time ranges with multivariate data, from reports to a structured tabular data into a scatterplot visualization. This implementation could be starting step for medical report information visualization.  Hence, bridge the gap of complex medical report and visual tool representation.  The evaluation can be done on a novel approach with different survey doctors and patients and try to check the criterion available on previously researched survey papers. This could assist in fitting in any changes required to the previous architecture or any added information details not identified or overlapped in the visualization tool.

In \cite{Blumenstein_undated-hv} evaluation of visualization was done on certain papers on \textit{infoViz} for different devices such as  projector, mobile phones and tablets for comparison. The comparison was performed on papers mainly extracted from IEEE and ACM libraries. The key criteria is based on the type of visualization on the devices, environment of usage, communication through visualization, data analysis, user performance and experience. From the evaluation, the survey derived that the target device or simulator information is missing for the majority of the papers. Secondly, the visualization tested on is a single device has missing information which makes it difficult to reproduce the research by others. User surveys done for the evaluation lack the versatility of users recorded. The chart for graph visualizations are mainly focused on scatter plots,  bar graphs, line graphs and histograms which might restrict the information to reflect on the visualizations. \cite{Mcnabb_undated-hx} discusses several information visualization strategies, with our focus on time-series data or visualization over higher time-slices. The survey provides insights into various visualization methods categorized and highlighted the value of empirical research on \textit{infoviz} modules. \cite{Wu2022-aj} derives modules of visualization with common interest from various fields. They build a framework of What-Why-How taxonomy where “What” results in ground truth of the visualization, “Why” for enhancement and generation of the visualization and “How” stands for processing and analysis of the results. This supports our selection of the appropriate visualization architecture for our model. This is precisely researched in \cite{Meyer2019-hg} which will give our research framework an efficient model with interactions on data, decision models that help us get the results we want, and reliable visualization tools. Specific \textit{infoviz} representation could not be applicable to other subjects due to the fact that knowledge is preliminary rather than conclusive and that the research is subjective. The final derivation from this survey paper concludes with visualization approach must acknowledge research features like informed, reflexive, abundant, plausible, resonant and transparent to achieve ideal design study.

Thus, we can conclude our \textit{infoViz} tool for mobile devices should prioritize these mentioned challenges to derive in the and build a diverse, robust and interactive cross platform visualization, with transparency on implementation and user information. Since, this is a novel approach, we try to derive the framework on the list of features from the research discussed.

\section{Framework}
During our exploration we found the above mentioned 16 papers can assist us in building the \textit{Infoviz} tool. Some of them are state-of-art visualisation tools and models. Thus we have identified key factors which are implemented to extract information, verify diverse platform, implement the visualization and compare the information. Please see Figure \ref{fig:Model}. We adopt the following as standard concepts to guide our proposed model. 
\begin{itemize}
	\item Semantics and data extraction
	\item Visualization models and patterns
	\item Data transformation
	\item Evaluations on efficacy of the model
\end{itemize}

This approach is modeled after a hierarchical task analysis approach and it constrains the information into a meaningful outcome as service to the visual tool framework. We discuss each of them in further details below.

\subsection{Semantics and data extraction} 
This is the preliminary stage of the model flow where data and information is extracted from the reports. The most challenging part of extracting a data is the diverse format available for input data. We have for now considered Portable Document Format (PDF) format to acknowledge the reports generally used and readily available. Our approach is derived as raw data is gathered from  standard multiple PDF report, extract the information, structure the details into a data table using data identification and structuring which will help in generating visual plot in different platform such as web, mobile and tablets. 

\begin{figure}[!ht]
  \begin{center}
  \includegraphics[scale=.05, width=100mm, height= 60mm]{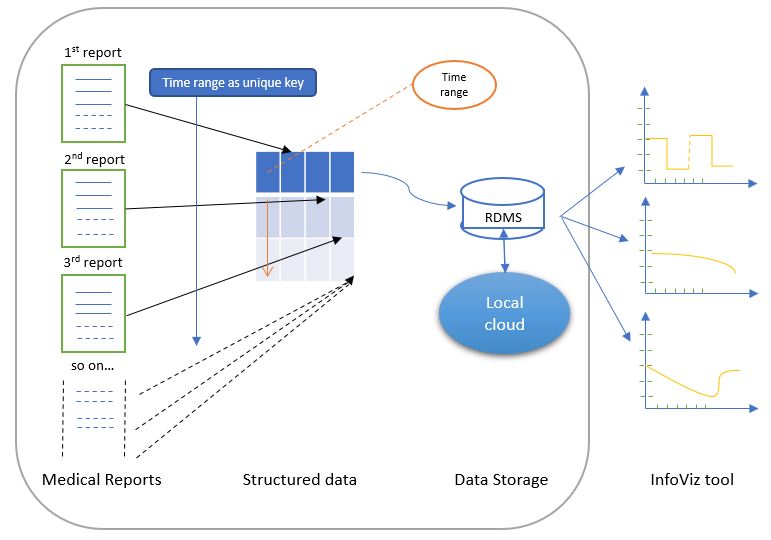}
  \caption{The figure for manipulating multiple reports to a structured data using time range as unique key between all the data semantics. Medical reports data are extracted with respect to date and time and send in a tabular form. In the next step, structured data uses a dynamic approach to align the data with varying column for every report. This data is then stored in relational database (RDMS) and connected to a local cloud. The local cloud is a storage which can be accessed from anywhere but the storage is a local drive for particular user.}
  \label{fig:DataExtraction}
  \end{center}
\end{figure}

The model organises the data from multiple report into a tabular format. To achieve this we consider the \textit{PDFDataExtractor} suggested in \cite{Zhu2022-ht}. In this extraction process focuses on quality data mining, selecting the important information. The report can be read and information are captured using name-entity-recognition to form a metadata file for a structured storage which could easily populate the table, identifying existing column or creating new column for the reports informatics. The primary link between the different report details is the time duration. 
This time range assists in binding the unstructured data into a relational data. The data interpretation into columns is a dynamic structure as shown in Figure \ref{fig:DataExtraction}. The number of structured table columns is directly influenced from the number of reports the patient has. The structured data can be stored in the relational database for further necessary assessment. The structured number of column can be represented by,
{\par\centering$f(r) = \sum_{k=1} ^{m} k r$\par where \(m\) is the number of extracted data from a single report, \(r\) is the number of reports required.}

\subsection{Visualization models and patterns} 
As per our discussion in previous section, the visualization should be able to accommodate multiple reports and seamlessly visualize the time ranges without compromising on the information extraction. For a more transparent visualization we considered two types: \textit{line graph} and \textit{radial graph} of visualization which could give us accurate representation of the complex report shown in Figure \ref{fig:DataExtraction}.

\begin{figure}[!ht]
  \centering
  \includegraphics[scale=.1, width=160mm, height= 100mm]{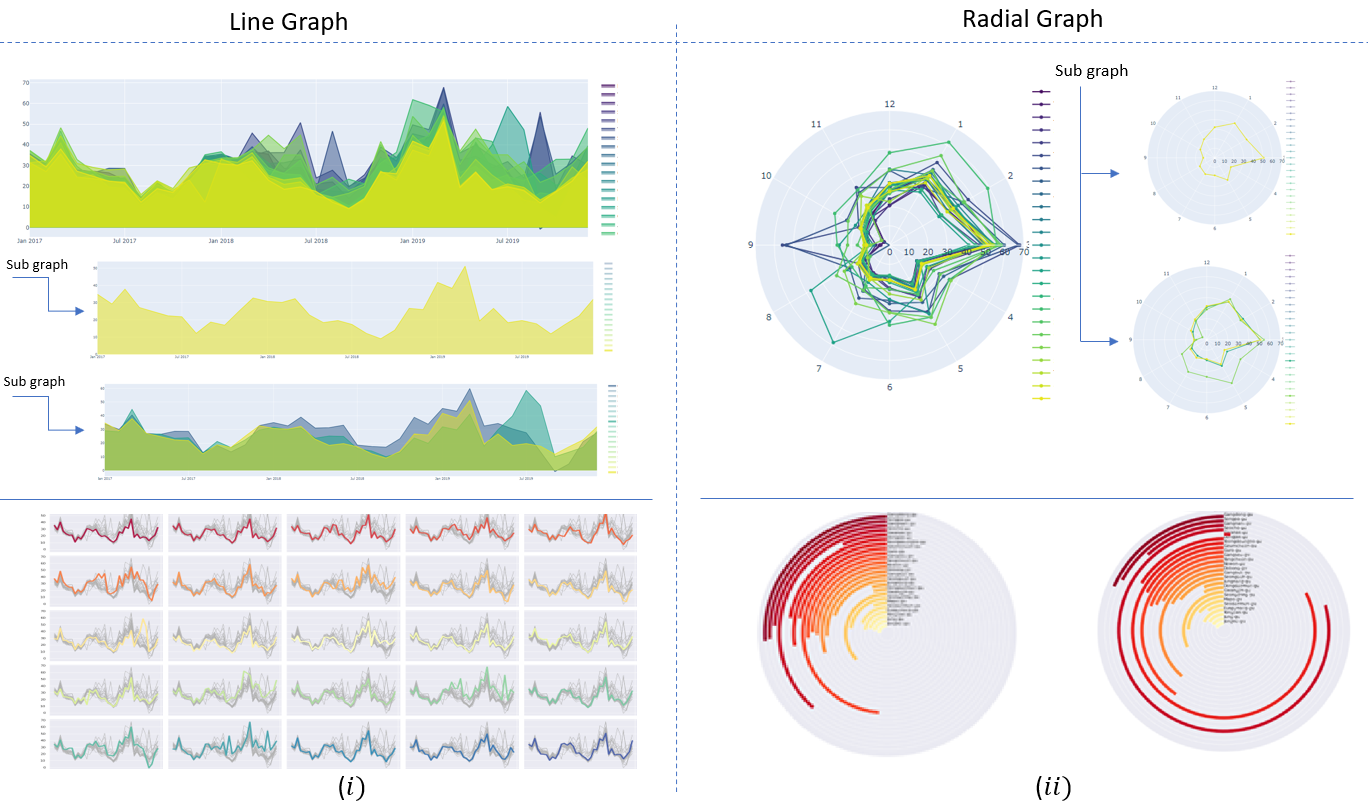}
  \caption{The \textit{Infoviz} for medical reports for multivariate system as well as longer time span can be considered using linear and radial graph plot. Figure 3-\textit{(i)} shows the line graph where time span is represented over \textit{x-axis} and \textit{y-axis} represents the frequency or volume of the multivariate data. In Figure 3-\textit{(ii)} we implement a radial graph representation. A radial graph's component renders data as a circle made up of a number of data points. The lower values are close to originating co-ordinates and the higher values are farther away along the periphery. Reference visualization from \cite{Boriharn2022-za}}
  \label{fig:Visualize}
\end{figure}
The visualization process is very unique process of extracting the maximum information with a visual glance. Therefore the information should follow some norms to make it readable for the user. As discussed in \cite{Midway2020-ko} the 10 steps should be addressed in \textit{Infoviz}.
In the first steps, recognising the data and representation int graph structure should be identified as we have done in Figure 3. This requires identifying the data which is already structured but need preprocessed before a visual graphics representation can be modeled. The preprocessing is required for labelling the graph onto the geometric shapes. The shapes can also be identified using different colors attracts more attention for the user. The data is normalized as per normal symptom level limits for comparative study on the graphs.

A line graph as shown in Figure 3-\textit{(i)} illustrates the connection between different data plots over a time range. Two points from different time scale are connected introducing a plot with visual graphic information of changing value. The line graph representation can be more accurately described as multiple line graph with compound line structure. A compound line structure checks for every individual line subdivided to extract more information from region of interest. A multiple line graph can be represented with values \textit{x} and  \textit{y} axis for every data point and connecting both points with the simple line equation \(y = mx + c\) , where \(x\) is the time range and \(y\) is the level or density f the concerned report value. Each pair of point is considered independent of the other points and points are considered start and endpoint respectively. For example, the line computed between \(\{x_1,y_1\}\) and \(\{x_2,y_2\}\) is completely independent from the line \(\{x_2,y_2\}\) and \(\{x_3,y_3\}\) with start and end point chosen between the points considered. For the radial graph, they can be represented by the top plot in Figure 3-\textit{(ii)} where the radii of the determines the value of the graph plot. This is similar to the line graph where each radii value is considered to be a data point and the line is plotted to connect the two adjacent points. The radial charts are important functionality to represent multivariate data for comparison  over temporal data. The dynamic behavior of the radial chart works on time ranges. The data points in the circular 2 dimensional plot increases with longer time spans. The below graph in Figure 3-\textit{(ii)} is a radial bar graph which is also a unique way to represent higher time ranges. This plot takes into consideration every column as consideration of a radii value and plots the column value on the plot. Radial bar graph is a efficient tool for comparison with specific columns easily distinguishable from other columns. In the next steps we concentrate on the facet and panel of the graphs which will describe our goal for mobile devices by data transformation from a relational data to a screen specific visual graph.

\subsection{Data transformation} 
The transformation of data is one of the complex characteristics on a  \textit{Infoviz} tool. If misrepresented on different screen size, we can have loss of critical data for analysis. Thus, we were encouraged to survey for detailed research done on the transformation.  Most of the tools are developed in a high end computers with large screens and high resolution. This makes the visualization difficult to adapt to smaller screens. Unknowingly many complexity rises for the transformation. \textit{MobileVizFixer} discussed in our literature survey is an ideal research for the transferring the visual data from a larger screen such as monitors to a portable smaller screens. Though the extraction process in the paper uses reinforcement learning algorithms, we tried to emulate similar issues addressed on the paper such as blank spaces, unreadable texts and fit to screen with resolution identification. In this proposed method we identify the the screen resolution and try to manage the data accordingly. As shown in Figure 4, the visual graphs when transformed into a smartphones \textit{infoviz} takes the lateral approach to accommodate all the necessary information values for higher time series. The same visualization when transformed into a bigger screen such as tablet the vertical view is encouraged for more information. 
\begin{figure}[!ht]
  \centering
  \includegraphics[scale=.1, width=100mm, height= 70mm]{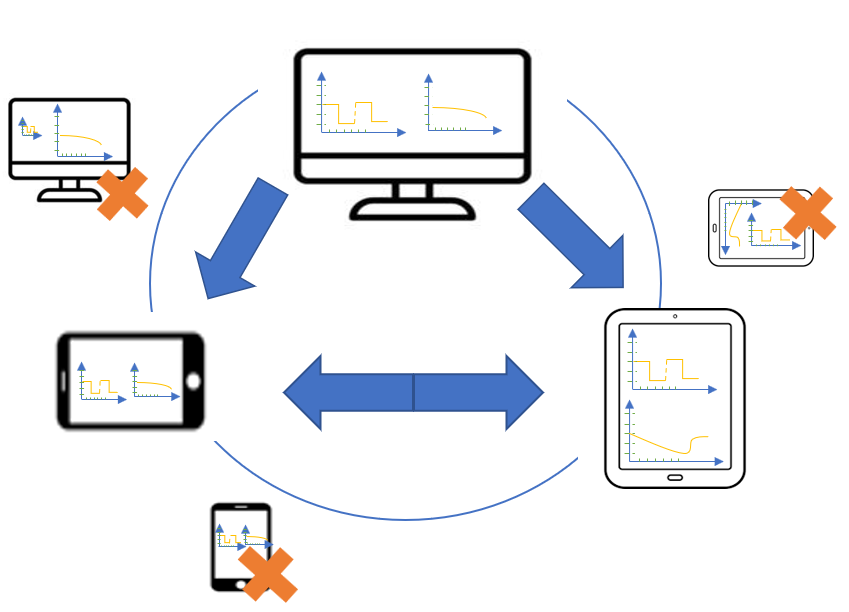}
  \caption{An illustration the data transformation and how the screen size effects the visualizing models. An accurate visualization modeling will be effective if the model adapts to the screen and hardware support of the \textit{Infoviz}. The figure suggests equal ratio, alignment of multivariate graph and fit to screen properties on monitors, tablet and smartphones respectively.}
  \label{fig:Visualize}
\end{figure}
Notably, the discarded views shown in Figure 4, for mobile and tablets illustrates the problems faced during the transformation. The graph having larger resolution will not fit to screen of a mobile and could easily hinder the visual informatics. Also, the lateral view and vertical view if not independently addressed could deform the visual representation. Thus, the screen identification and visual manipulation is a important part of the information visualization for multivariate data with higher time series.

\subsection{Evaluations on efficacy of the model}
 \begin{table}[ht]
\centering
\begin{tabular}{ |c|c|c } 
\hline
Features & Results  \\
\hline
\hline
Multivariate data accommodation & Yes \\
\hline
Higher time series graph & Yes  \\ 
\hline
Device transparency & Yes \\ 
\hline
Descriptive details on implementation & Yes \\ 
\hline
Dynamic data accumulation & Yes \\ 
\hline
\end{tabular}

\caption{Features provided on the framework for successful implementation}
\label{Table:1}
\end{table}

Evaluation of the efficiency of the model describes our framework's capability can be used in building a data driven \textit{Infoviz} tool to personalize and visualize the medical reports for complex chronic diseases. The key features addressed in our model are multivariate data and time series representation on the visualization. Along with this, we have also addressed the data transparency to other devices. The model will work on other different resolution and dynamic data allocation process. The modeling has enabled us to look into most asked queries which are suggested in the survey papers in previous section. The data representation from the ground truth with higher transparency on the model description. Our each module in the framework work on heuristic way to accommodate the complexity of \textit{Infoviz} tool. 

The multivariate data describes multiple report structuring, time series depends on the high time range which is a important factor for chronic diseases. Device transparency can be addressed in versatility of screen resolution and details of implementation can be described in the model shown in Figure \ref{fig:Model}. And lastly, dynamic data is an example of building dynamic columns on the data extraction to fit in more reports on the structured data. Henceforth we will try to implement a \textit{Infoviz} tool with this approach and derive the conclusion or any changes required into the defined framework.

\section{Conclusion}

For patients, healthcare professionals and their support network, visualizing the medical histories of patients with complicated chronic conditions is a complicated activity.
Patients with DCC's may need to receive therapy for a number of, frequently unrelated chronic conditions at the same time and may also be given conflicting treatment suggestions.
We provide a framework by thorough consideration to visualize the medical reports from symptoms, and treatment suggestions provided by DCC's data in order to decrease therapy conflicts and enhance patient quality of life and care. We investigate numerous visualization approaches and paradigms in this work and examine the use of various models and paradigms to display complex medical data.
In a single graphic model, we suggest a method for converting the unstructured input into temporal slices. We also describe the method we used to carefully transfer complex DCC records into hardware devices like structured data tables and graphics.

In future, we would like to implement the framework in a tool using the detailed summary given in the previous section. The sub goals from data extraction to visualization on multiple devices, to evaluation of model will be considered astutely to successfully implement the \textit{Infoviz} tool. Finally, we will evaluate the accuracy of the framework through this tool by human subjects for feedback. This approach can be considered as a benchmark for our further development on visualization in medical field research. 

\bibliographystyle{unsrtnat}
\bibliography{references}  






\end{document}